\documentclass{ws-procs9x6}

\def\beq{\begin{equation}}
\def\eeq{\end{equation}}
\def\beqa{\begin{eqnarray}}
\def\eeqa{\end{eqnarray}}

\def\za{\alpha}
\def\zb{\beta}

\def\lsim{\mathrel{\raise.3ex\hbox{$<$\kern-.75em\lower1ex\hbox{$\sim$}}} }
\def\gsim{\mathrel{\raise.3ex\hbox{$>$\kern-.75em\lower1ex\hbox{$\sim$}}} }

\begin{document}
\thispagestyle{empty}

\onecolumn

\begin{flushright}
NCU-HEP-k016  \\
Apr 2004
\end{flushright}

\vspace*{.5in}

\begin{center}
{\bf  \boldmath \protect \Large On Extended Electroweak Symmetries.  }\\
\vspace*{.5in}
{\bf  Otto C.W. Kong}\\[.05in]
{\it Department of Physics, National Central University, \\ Chung-li, TAIWAN 32054 \\
E-mail: otto@phy.ncu.edu.tw}

\vspace*{1.in}
\end{center}
\abstracts{We discuss extensions of the Standard Model through
extending the electroweak gauge symmetry. An extended electroweak
symmetry requires a list of extra fermionic and scalar states. The former is
necessary to maintain cancellation of gauge anomalies, and largely fixed
by the symmetry embedding itself. The latter is usually considered quite
arbitrary, so long as a vacuum structure admitting the symmetry breaking
is allowed. Anomaly cancellation may be used to link the three families
of quarks and leptons together, given a perspective on flavor physics.
It is illustrated lately that the kind of models may also have the so-called
little Higgs mechanism incorporated. This more or less fixes the scalar
sector and take care of the hierarchy problem, making such models of
extended electroweak symmetries quite appealing candidates as TeV
scale effective field theories.
}

\vfill
\noindent --------------- \\
$^\star$ Talk presented  at CGC 2003 (Dec 4-17-21), Fort Lauderdale, FL USA\\
 --- submission for the proceedings.  
 
\clearpage
\addtocounter{page}{-1}

\title{\bf\boldmath\protect On Extended Electroweak Symmetries}

\author{OTTO~C.~W. KONG\footnote{\uppercase{W}ork partially
supported by grant \uppercase{NSC} 92-2112-\uppercase{M}-008-044 of the 
\uppercase{N}ational \uppercase{S}cience \uppercase{C}ouncil of \uppercase{T}aiwan.}}

\address{Department of Physics, National Central University,
Chung-li, TAIWAN 32054\\ 
E-mail:  otto@phy.ncu.edu.tw}

\maketitle

\abstracts{We discuss extensions of the Standard Model through
extending the electroweak gauge symmetry. An extended electroweak
symmetry requires a list of extra fermionic and scalar states. The former is
necessary to maintain cancellation of gauge anomalies, and largely fixed
by the symmetry embedding itself. The latter is usually considered quite
arbitrary, so long as a vacuum structure admitting the symmetry breaking
is allowed. Anomaly cancellation may be used to link the three families
of quarks and leptons together, given a perspective on flavor physics.
It is illustrated lately that the kind of models may also have the so-called
little Higgs mechanism incorporated. This more or less fixes the scalar
sector and take care of the hierarchy problem, making such models of
extended electroweak symmetries quite appealing candidates as TeV
scale effective field theories.}

\begin{center}
\em Dedicated to Paul Frampton
\end{center}

\section{Introduction}
This talk is my contribution to the event celebrating the 60th birthday of Paul
Frampton. The subject here is extending electroweak symmetries, in particular,
as an approach to particle physics beyond the Standard Model (SM). The focus 
is on my own works on the subject, which began during the time I was a student
studying under Paul's supervision. I am getting back to the topic lately, with
some studies more related to some of Paul's own works but in the name
of little Higgs.

The SM is a model of interactions dictated by an
$SU(3)_C\times SU(2)_L\times U(1)_Y$ gauge symmetry, with a anomaly
free chiral fermion spectrum and a Higgs multiplet responsible for the
spontaneous breaking of the electroweak (EW) symmetry $SU(2)_L\times U(1)_Y$.
Extending the EW gauge symmetry extends the SM while adding new
fermions and scalars. This is to be contrasted with other approaches such
as grand unification and/or supersymmetry. Compares with the latter approaches,
extending EW symmetry may look like less popular or not so well motivated.
Grand unification aims at providing a unified picture to all the otherwise separate
parts of the gauge symmetry and their independent couplings, though only at a scale 
of about $10^{-16}$ GeV.  All that can be achieved, in the case of $SU(5)$, without 
the need for extra fermionic states. Supersymmetry used the beautiful boson-fermion
symmetry to tackle the hierarchy problem, essentially extending the chiral
nature of the fermions to fix the problem for the scalar sector. Putting the two
together provides a theoretical structure that promises to ``explain" more
or less all of particle physics. However, the large extrapolation over the
many order of magnitudes of particle physics desert may certainly be
taken with suspicion. Moreover, the approaches do not provide any new
insight into the difficult problem of the origin of flavor structure. Why there are 
three families of SM fermions is still a fundamental problem that we have 
no credible approach to handle. On the contrary, extending the EW symmetry
may provide some new perspectives to the flavor problem. It can even provide
an alternative solution to the hierarchy problem, in the name of the so-called
little Higgs mechanism\cite{acg}.

\section{Looking at the Fermionic Spectra}
The spectrum of SM fermion in one family is like perfection, essentially dictated by 
gauge anomaly cancellation conditions.  To illustrate the point of view, we recall 
our earlier argument\cite{unc67}. Assuming that there exist a minimal multiplet 
carrying nontrivial quantum numbers of each of the component gauge groups, one 
can obtain the one-family SM spectrum as the unique solution by asking for the 
minimal consistent set of chiral states. Consistence here refers to the perfect 
cancellation of nonvanishing contributions to various gauge anomalies from 
individual fermionic states. A vectorlike set (or pair) is trivial but not  interesting. 
Only chiral states are protected from heavy gauge invariant masses and relevant
to physics at the relatively low energy scale. 

The above suggested derivation of the one-family SM spectrum goes as follow.
We are essentially starting with a quark doublet,  with arbitrary hypercharge
normalization. The two $SU(3)_C$ triplets require two antitriplets to cancel
the anomaly. Insisting on the chiral spectrum means taking two quark singlets
here, with hypercharges still to be specified. Now, $SU(2)_L$ is real, but has
a global anomaly. Cancellation requires an even number of doublets, so at least
one more beyond the three colored components in the quark doublet. There
are still four anomaly cancellation conditions to take care of. They are the
$[SU(3)_C]^2 U(1)_Y$, $[SU(2)_L]^2 U(1)_Y$, $[grav]^2 U(1)_Y$, and
$[U(1)_Y]^3$. We are however left with three relative hypercharges to fit the
four equations, actually without a possible solution. A rescue comes from simply
adding a $U(1)_Y$-charged singlet. But the four equation for four unknown
setting is misleading. The $[U(1)_Y]^3$ anomaly cancellation equation is cubic
in all the charges, with no rational solution guaranteed. The SM solution may
actually be considered a beautiful surprise. Moreover, the perspective may 
be the best we have on understanding {\it why there is what there is}. 

\begin{table}[b]
\noindent\hrule 
\vspace*{.2in}
\begin{center}
\footnotesize
\begin{tabular}{|c|c|r|r|r|r|r|cc|}
\hline\hline
 multiplets & $X$ &
\multicolumn{5}{|c|}{Gauge anomalies} &  \multicolumn{2}{|c|}{$U(1)_Y$ states } \\ 
\hline 
&  &	trX & 44X	& 33X &  22X &	{$X^3$} &				& \\
\hline								  		
${\bf (4,3,2)}$		&	{\bf 1}  &	24  &	6  &	8 &	12 &	24	& 	3\ {\bf 1}($Q$)	   	& {\bf -5}($Q^{'}$) \\
${\bf (\bar{4},\bar{3},1)}$ &	{\bf 5}  &	60  &	15 & 	20 &	 &	1500	& 	3\ {\bf -4}($\bar{u}$)	& {\bf 2}($\bar{d}$) \\
${\bf (\bar{4},1,2)}$	&	{\bf 3}  &	24  &	6  &	&	12 &	216	& 	3\ {\bf -3}($L$)	& {\bf 3}($\bar{L}$) \\
${\bf (\bar{4},1,1)}$	&	{\bf 9}  &	36  &	9  &	&	 &	2916	& 	3\ {\bf -6}($\bar{E}$)	& {\bf 0}($N$) \\
${\bf (6,1,1)}$		&	{\bf -18} &	-108 &	-36 &	&	 &	-34992	& 	3\ {\bf 6}($E$)	& 3\ {\bf 12}($S$) \\ \hline
${\bf (1,\bar{3},2)}$	&	{\bf -10} &	-60 &	&	-20 &	-30 &	-6000	& 	\multicolumn{2}{|c|}{{\bf 5}($\bar{Q}^{'}$)} \\
${\bf (1,\bar{3},1)}$	&	{\bf -4} & 	-12 &	&	-4  &	 &	-192	& 	\multicolumn{2}{|c|}{{\bf 2}($\bar{d}$)} \\
${\bf (1,\bar{3},1)}$	&	{\bf -4} &	-12 &	&	-4  &	 &	-192	& 	\multicolumn{2}{|c|}{{\bf 2}($\bar{d}$)} \\ \hline
${\bf (1,1,2)}$		&	{\bf 6}  &	12  &	&	&	6  &	432	&	\multicolumn{2}{|c|}{{\bf -3}($L$)} \\
$3\ {\bf (1,1,1)}$ 	&	{\bf 24} &	72  &	&	&	&	41472	&	\multicolumn{2}{|c|}{3\ {\bf -12}($\bar{S}$)} \\ 
$3\ {\bf (1,1,1)}$ 	&	{\bf -12} &	-36 &	&	&	&	-5184	&	\multicolumn{2}{|c|}{3\ {\bf 6}($E$)} \\	\hline
\multicolumn{2}{|r|}{\it Total}    	&	0   &	0  &	0  &	0  & 	0	& 				& \\		
\hline\hline
\end{tabular}
\end{center}
{\footnotesize Table I. An $SU(4)_A\times SU(3)_C\times SU(2)_L\times U(1)_X$ 
spectrum embedding three SM families.}
\end{table}

\normalsize
We would also like to take the opportunity here to briefly sketch the next step taken
in Ref.\cite{unc67}, to further illustrate our perspective. The results there also may
be considered a worthy comparison with our little Higgs motivated flavor/family
spectrum presented below, from the point of view of the origin of the three families.
The major goal of Ref.\cite{unc67} is to use a similar structure with an extended
symmetry to obtain the three families. For example, one can start with some
$SU(4)\times SU(3)\times SU(2)\times U(1)$ gauge symmetry and try to obtain the 
minimal chiral spectrum contain a {\boldmath $(4,3,2)$} multiplet --- the simplest one
with nontrivial quantum number under all component groups. Having a consistent
solution is not enough though. In order for the spectrum be of interest, we ask the 
spectrum to yield  the chiral spectrum of three SM families plus a set of vectorlike
states under a feasible spontaneous symmetry breaking scenario, {\it i.e.} when
the gauge symmetry is broken to that of the SM. Ref.\cite{unc67} has only
partial success. A consistent group theoretical SM embedding could be obtained
but only with a slight addition to the minimal chiral spectrum obtained from anomaly
cancellation considerations alone. We give an example in Table I.

Next, we recall the fermionic spectrum from a simple model of extended EW
symmetry, the 331 model from Paul himself\cite{331}. The model has the EW
symmetry extended to an $SU(3)_L\times U(1)_X$. To have a consistent spectrum
of chiral fermions, one may first look into how the SM doublets are to be embedded
into multiplets of $SU(3)_L$. It is interesting to note here that a naive family
universal embedding would not work. The $SU(3)_L$ anomaly would not cancel.
Instead, the model has the $(t,b)$ doublet embedded into a {\boldmath$\bar{3}$} while
the quark doublets of the first two families into {\boldmath$3$}'s, with all leptonic 
doublets embedded into {\boldmath$\bar{3}$}'s. The fact that the number of color equals the 
number of families makes the anomaly cancellation possible. All extra quark states here 
are exotic, with charges $\frac{5}{3}$ and $\frac{-4}{3}$. There are no extra leptonic 
states though. The 331 model spectrum is given in Table II.

\begin{table}[h]
\noindent\hrule \small
\vspace*{.2in}
\begin{center}
\begin{tabular}{|c|cc|}
\hline\hline
				& \multicolumn{2}{|c|}{$U(1)_Y$-states}			\\  \hline
{${\bf (3_{\scriptscriptstyle C},\bar{3}_{\scriptscriptstyle L} ,\frac{2}{3})}$}                 &  ${\bf \frac{1}{6}}$[$Q$]               & ${\bf \frac{5}{3}}$($T$)       	\\
{2\ ${\bf ({3}_{\scriptscriptstyle C},{3}_{\scriptscriptstyle L} ,\frac{-1}{3})}$}     &  2\ ${\bf \frac{1}{6}}$[2\ $Q$]               & 2\ ${\bf \frac{-4}{3}}$($D,S$)     \\
{$3\ {\bf (l_{\scriptscriptstyle C} ,\bar{3}_{\scriptscriptstyle L} ,0)}$}           & 3\  ${\bf \frac{-1}{2}}$[3\ $L$]          	& 3\ {\bf 0}(3\ $E^+$)   \\
$3\ {\bf (\bar{3}_{\scriptscriptstyle C},1_{\scriptscriptstyle L} ,\frac{-2}{3})}$        & \multicolumn{2}{|c|}{4\ ${\bf \frac{-2}{3}}$ ($\bar{u}, \bar{c}, \bar{t}$)} 	     	\\
$3\ {\bf (\bar{3}_{\scriptscriptstyle C},1_{\scriptscriptstyle L} ,\frac{1}{3})}$        & \multicolumn{2}{|c|}{5\ ${\bf \frac{1}{3}}$ ($\bar{d}, \bar{s}, \bar{b}$)} 	                 	\\
$1\ {\bf (\bar{3}_{\scriptscriptstyle C},1_{\scriptscriptstyle L} ,\frac{-5}{3})}$              &    \multicolumn{2}{|c|}{ ${\bf \frac{-5}{3}}$  ($\bar{T}$)   }      	\\
$1\ {\bf (\bar{3}_{\scriptscriptstyle C},1_{\scriptscriptstyle L} ,\frac{4}{3})}$              &    \multicolumn{2}{|c|}{2\ ${\bf \frac{4}{3}}$  ($\bar{D}, \bar{S}$)   }      	\\
\hline\hline
\end{tabular}
\\[.2in]
{\small Table II. A (331) model spectrum for $SU(3)_L\times U(1)_X$ extended EW symmetry.}\\[.1in]
\end{center} \hrule
\normalsize
\end{table}

\section{Extended EW Symmetries of $SU(N)_L\times U(1)_X$}
Looking at the model spectrum of Table II, one may wonder if the construction is in any 
sense unique, and if similar anomaly free spectra exist for a different extended EW
symmetry. We look into the question lately and have the general solution. It turns out
quite simple and straightforward.

For an extended EW symmetry of $SU(N)_L\times U(1)_X$, the SM doublets may be
embedded into {\boldmath $N$}'s or {\boldmath $\bar{N}$}'s. Embedding one 
quark doublet into an {\boldmath $N$} and the two others into {\boldmath $\bar{N}$}'s
while putting all lepton doublets into {\boldmath $N$}'s does give a prescription with
canceled  $SU(N)$ anomaly. The a bit of surprising part is that no matter how one 
chooses to embed $U(1)_Y$ into $SU(N)_L\times U(1)_X$, simply completing the list of 
chiral states with appropriate $SU(N)$ singlets to ensure vectorlike matchings at the
QCD and QED level does yield a completely anomaly free spectrum, essential unique
for the particular symmetry embedding. The number of possible consistent model 
spectra of the type is then equivalent to the number of admissible symmetry embeddings.
The latter can conveniently be parametrized by the choice of electric charges for the
extra $N-2$ quark states sharing the  {\boldmath $N$} multiplet with the $(t,b)$
doublet\cite{010}. We have no room in this write-up to elaborate on the details though.

\section{Little Higgs and Extended Electroweak Symmetries}
The little Higgs mechanism\cite{acg} has been proposed as new solution to the hierarchy
problem. More precisely, it alleviates the quadratic divergent quantum correction to the
SM Higgs states and admits a natural little hierarchy between the EW scale and a higher
scale of so-called UV-completion at around 10 TeV above which further structure would be
hidden. The idea is a rather humble bottom-up approach then; but experimental hints
at the existence of such a little hierarchy has been discussed\cite{Bar}.
What is relevant for our present discussion is that a little Higgs model necessarily
has an extended gauge symmetry, EW or beyond, and extra fermion(s). The latter includes
a heavy top $T$ quark.

Simple little Higgs model(s) based on an extended EW symmetry has been introduced
by  Kaplan and Schmaltz\cite{KS}, though the authors failed to properly address the
structure of the fermionic sector. The gauge symmetry considered are 
$SU(3)_L\times U(1)_X$ and $SU(4)_L\times U(1)_X$. We discuss completion of the 
kind of models with consistent, anomaly free, fermionic spectra and the resulted 
implications on the flavor structure of the models in Ref.\cite{010,009,012}.  Naively,
so long as one pick a model spectrum with an extra $T$ quark in the $(t,b)$ containing
{\boldmath $N$} multiplet (here $N=3$ or $N=4$, for example), one have potentially
a extended EW little Higgs model. The $T$ quark may be used to cancel the quadratic
divergent contributions (only at 1-loop level) to the SM Higgs mass from the $t$ quark, 
while the extra EW gauge bosons to do the same for their SM counterparts. The 
scalar/Higgs sector has to be explicitly constructed though, to have the SM Higgs doublet 
embedded as (pseudo-)Nambu-Goldstone states of some global symmetry. It is an 
$[SU(3)]^2/[SU(2)]^2$ symmetry for the $SU(3)_L\times U(1)_X$ case, for instance. The 
Higgs sector symmetry is to be explicitly violated beyond the sector, in the gauge and 
Yukawa couplings of the Higgs multiplets. Such a scheme can be easily achieved with 
pair(s) of Higgs multiplets having the right quantum number to couple to the $(T,t,b,..)$
multiplet and a right-handed  $T$ singlet. However, there is source of further complication, 
related to the construction of a proper Higgs quartic coupling term\cite{KS}. We admit that, 
in general, the latter issue still have to be studied more carefully. We do have a definitely
complete and consistent model though. This is given by the fermion spectrum of Table
III, with the Higgs sector as given in Ref.\cite{KS}. Here below, we will focus on the
fermionic sector and flavor physics structure.

\begin{table}[b]
\hrule\vspace*{.2in}
\begin{center}\footnotesize
\begin{tabular}{|c|r|r|r|r|r|cc|}
\hline\hline
 & \multicolumn{5}{|c|}{Gauge anomalies} &  \multicolumn{2}{|c|}{$U(1)_Y$ states } \\ 
\hline 
    &  	 ${\rm tr}X$ & $LLL$	& $LLX$ &  $CCX$ &	$\!\!\!\!${\tiny(144)}$X^3$ &			& \\
\hline								  		
${\bf (3_{\scriptscriptstyle C},4_{\scriptscriptstyle L} ,\frac{5}{12})}$
      &    $5$ &	$3$  &	 ${5/4}$ &	${5/3}$ &  ${125}$	 &  ${\bf \frac{1}{6}}$[$Q$]               & 2\ ${\bf \frac{2}{3}}$($T$)    \\
2\ ${\bf ({3}_{\scriptscriptstyle C},\bar{4}_{\scriptscriptstyle L} ,\frac{-1}{12})}$
     &       $-2$ &	${-6}$ & 	${-1/2}$	 & ${-2/3}$ &	$-2$    &  2\ ${\bf \frac{1}{6}}$[$Q$]               & $\!\!\!\!$4\ ${\bf \frac{-1}{3}}$($2\ D,2\ S$) 	     \\
$3\ {\bf (l_{\scriptscriptstyle C} ,4_{\scriptscriptstyle L} ,\frac{-1}{4})}$
    &	 $- 3$&	$3$  &	   ${-3/4}$ &	 & ${-27}$	           & 3\  ${\bf \frac{-1}{2}}$[$L$]          	& 6\ {\bf 0}($N$)  	  \\
$5\ {\bf (\bar{3}_{\scriptscriptstyle C},1_{\scriptscriptstyle L} ,\frac{-2}{3})}$
   &	$\!\!$$-10$&	 &	&	 $\!\!$$-10/3$ & ${-640}$        & \multicolumn{2}{|c|}{5\ ${\bf \frac{-2}{3}}$ ($\bar{u}, \bar{c}, \bar{t},  2\ \bar{T}$)} 		\\
$7\ {\bf (\bar{3}_{\scriptscriptstyle C},1_{\scriptscriptstyle L} ,\frac{1}{3})}$  
   &	$7$&	&	&	 $7/3$ & ${112}$        & \multicolumn{2}{|c|}{7\ ${\bf \frac{1}{3}}$ ($\bar{d}, \bar{s}, \bar{b}, 2\ \bar{D},2\  \bar{S}$)}	\\
$3\ {\bf (1_{\scriptscriptstyle C},1_{\scriptscriptstyle L} ,1)}$   	
    &     $3$  &	 &	&	 &	$432$           &    \multicolumn{2}{|c|}{3\ ${\bf 1}$  ($e^+, \mu^+, \tau^+$)   }       	\\ 
 \hline
\multicolumn{1}{|r|}{ Total}    	&	0  &   0  &	0  &	0  & 	0	& 				& 	\\	
\hline\hline
\end{tabular}

\vspace*{.1in}
 {Table III.  The $SU(3)_C\times SU(4)_L\times U(1)_X$ chiral fermionic spectrum 
completing \\ the Kaplan-Schmaltz little Higgs model, with anomaly cancellation 
illustrated.}
\end{center}
\normalsize
\end{table}

With a specific choice of the extended EW symmetry, a little Higgs model can be built only 
with the inclusion of the $T$ quark state. For the $N=3$ case, that fixes the hypercharge
embedding and hence, from our anomaly cancellation study, the unique fermionic spectrum.
The spectrum can be read off from Table III, with only one set of the duplicated $T$, $D$,
$S$, and three $N$ states. Note that the $X$-charges will have to change accordingly.
For the $N=4$ case, one may consider variations of the model spectrum, essentially
by choosing a different set of states beyond that of the $N=3$ content. In particular,
a spectrum with a full set of duplicated, heavy, SM fermions look very interesting\cite{010}.
However, the scalar/Higgs sector has to be explicitly constructed then. Following 
exactly the construction of Ref.\cite{KS}, one may be restricted to the spectrum of
Table III, with trivial generalization to $N>4$ spectra extending on the content. At the 
moment, one sees no motivation to go for $N>4$ .

\section{Some Implications to Flavor Physics}
Unlike generic models of extended EW symmetries, we do not have much freedom
in picking a set of scalar multiplets with VEVs according to what mass generating
Yukawa couplings we may want to include. However, a careful checking of the Higgs
multiplets shows that phenomenologically acceptable  mass terms for the fermions, SM 
ones or heavy quarks, can be obtained for the explicit models 
discussed above. Here, we use the $SU(3)_L\times U(1)_X$ case for the demonstration,
in favor or simpler notation and expressions.

As touched on above, the little Higgs mechanism is to be implemented with two 
scalar multiplets having the right quantum number to couple to the chiral parts of
the $T$ quark. They are denoted by $\Phi_{\!\scriptscriptstyle 1}$ and 
$\Phi_{\!\scriptscriptstyle 2}$ below in the expression of which we give the Yukawa
part of the Lagrangian. The latter is constructed simply by tracing the quantum 
numbers and admitting all terms compatible with the gauge symmetries.
\beqa 
{\mathcal L}_{\rm\tiny Yukawa}  &=&   \lambda^t_{\!\scriptscriptstyle 1}\,\bar{t'}\,
\Phi_{\!\scriptscriptstyle 1} \, Q   +    \lambda^t_{\!\scriptscriptstyle 2}\,\bar{T'}\,
\Phi_{\!\scriptscriptstyle 2} \, Q
+\frac{1}{M} \lambda^u_{\za j} \, \bar{u'}_\za \,
\Phi_{\!\scriptscriptstyle 1} \, \Phi_{\!\scriptscriptstyle 2} \,  Q'_j 
\nonumber \\  && \quad +
 \lambda^{d1}_{\zb j} \, \bar{d'}_\zb \,
\Phi_{\!\scriptscriptstyle 1}^\dag \,   Q'_j  + \lambda^{d2}_{\zb j} \, \bar{d'}_\zb \,
\Phi_{\!\scriptscriptstyle 2}^\dag \,  Q'_j   + \frac{1}{M} \lambda^b_{\zb} \, \bar{d'}_\zb \,
\Phi_{\!\scriptscriptstyle 1} ^\dag\, \Phi_{\!\scriptscriptstyle 2} ^\dag\,  Q  \;,
\eeqa
where   $Q$ and $Q'_j$ denote (contrary to notation in Table III) the color triplet and antitriplets.
Note that we have to include dimension five terms here. Recall that the little Higgs model
actually has a high energy cut-off of only around a 10 TeV scale. 
The next step is to use the nonlinear sigma model expansion of the scalar multiplets in 
terms of the pseudo-Nambu-Goldstone states, which include the SM Higgs doublet 
$h$\cite{KS,009}. We recover
\beqa
&& \!\!\!
 {\mathcal L}_{\rm\tiny Yukawa} =
f\,(\lambda^t_{\!\scriptscriptstyle 1}\,\bar{t'}+  \lambda^t_{\!\scriptscriptstyle 2}\,\bar{T'})\, T
+ f\, ( \lambda^{d1}_{\zb j} \, \bar{d'}_\zb +   \lambda^{d2}_{\zb j} \, \bar{d'}_\zb) \, D_j
\nonumber \\  
&& \quad\quad
+ \frac{i}{\sqrt{2}} \, (\lambda^t_{\!\scriptscriptstyle 1}\,\bar{t}'
-  \lambda^t_{\!\scriptscriptstyle 2}\,\bar{T}')\,h
\left( \begin{array}{c} t \\ b \end{array} \right) 
- \frac{i \, \sqrt{2} \, f}{M} \lambda^u_{\za j} \, \bar{u'}_\za \, h \, 
\left( \begin{array}{c} u_j \\ d_j \end{array} \right) 
\nonumber \\  
&& \quad\quad
- \frac{i}{\sqrt{2}} \, (\lambda^{d1}_{\zb j} \, \bar{d'}_\zb 
-   \lambda^{d2}_{\zb j} \, \bar{d'}_\zb)\, h^\dag   
\left(   \begin{array}{c} u_j \\ d_j \end{array}    \right) 
+  \frac{i \, \sqrt{2} \, f}{M} \lambda^b_{\zb} \, \bar{d'}_\zb  \, h^\dag \, 
\left( \begin{array}{c} t \\ b\end{array}   \right) + \cdots 
 \label{Ld} \eeqa
The expression shows that all the heavy quark state, $T$, and $D_j$ (or $D$ and $S$)
get Dirac mass at scale $f$ of the VEVs of $\Phi_{\!\scriptscriptstyle 1}$ and 
$\Phi_{\!\scriptscriptstyle 2}$, and standard Yukawa couplings for the SM quarks
and Higgs doublet are all available. However, the expression also indicates that one has to
expect mass mixings among heavy and SM quark states. The nature of the extra heavy 
quarks and their mass mixings with the SM counterparts dictate stringent constraints 
on the related couplings and interesting flavor physics.

\section{Conclusions}
The bottom line here is that sensible discussion of flavor physics of a little Higgs model
is not possible before the full fermion spectrum is spelt out. The latter is constrained
by gauge anomaly cancellation. We exhibit at least one complete model here on which
detailed flavor physics still have to be studied. For the kind of models, the fermionic part
has a family non-universal flavor structure just like that of the 331 model, linking
the three SM families into one fully connected set. 
Gauge anomaly cancellation should play a major role on constructing the fermionic 
completion of any little Higgs model. This is, unfortunately, an issue that has been largely 
overlooked in the literature.

In summary, we see that studies of extended EW symmetries has arrived at the point
of furnishing all round models of beyond SM physics addressing more or less all the
concerns of particle physics, including the hierarchy problem. Such a model then
has almost no arbitrary parts to be chosen at model-builders' discretion. It has generic
appeals, but are also very humble, liable to various stringent precision EW and flavor 
physics constraints and begs UV-completion about an order of magnitude in energy scale 
above that of the electroweak theory. Building models of the kind, and studying
their phenomenology in details, as well as checking the predictions experimentally 
should be a worthy endeavor.

\input{016+}
\end{document}